\documentclass[twocolumn,showpacs,preprintnumbers,amsmath,amssymb]{revtex4}
\usepackage{graphicx}% Include figure files
\usepackage{dcolumn}% Align table columns on decimal point
\usepackage{bm}% bold math
\def\be{\begin{equation}}
\def\ee{\end{equation}}
\def\bea{\begin{eqnarray}}
\def\eea{\end{eqnarray}}

\newcolumntype{d}{D{.}{.}{3.1}}

\begin{document}

\preprint{APS/123-QED}

\title
{Multinucleon transfer reactions in closed-shell nuclei 
}

\bigskip

\author{S.~Szilner$^{1,4}$}
\author{C.~A.~Ur$^{2,5}$}
\author{L.~Corradi$^1$}
\author{N.~M\u{a}rginean$^1$}
\author{G.~Pollarolo$^3$}
\author{A.~M.~Stefanini$^1$}
\author{S.~Beghini$^2$}
\author{B.~R.~Behera$^{1}$} 
\author{E.~Fioretto$^1$}
\author{A.~Gadea$^1$}
\author{B.~Guiot$^1$}
\author{A.~Latina$^1$}
\author{P.~Mason$^2$}
\author{G.~Montagnoli$^2$}
\author{F.~Scarlassara$^2$}
\author{M.~Trotta$^7$}
\author{G.~de~Angelis$^1$}
\author{F.~Della~Vedova$^2$}
\author{E.~Farnea$^2$}
\author{F.~Haas$^6$}
\author{S.~Lenzi$^2$}
\author{S.~Lunardi$^2$}
\author{R.~M\u{a}rginean$^2$}
\author{R.~Menegazzo$^2$}
\author{D.~R.~Napoli$^1$}
\author{M.~Nespolo$^2$}
\author{I.~V.~Pokrovsky$^1$}
\author{F.~Recchia$^2$}
\author{M.~Romoli$^7$}
\author{M.-D.~Salsac$^6$}
\author{N.~Soi\'{c}$^4$}
\author{J.~J.~Valiente-Dob\'{o}n$^1$}

\affiliation
{$^1$ Istituto Nazionale di Fisica Nucleare, Laboratori Nazionali di Legnaro,
I-35020 Legnaro, Italy}

\affiliation
{$^2$ Dipartimento di Fisica, Universit\`{a} di Padova, and Istituto Nazionale
di Fisica Nucleare, I-35131 Padova, Italy}

\affiliation
{$^3$ Dipartimento di Fisica Teorica, Universit\`a di Torino, and
Istituto Nazionale di Fisica Nucleare, 10125 Torino, Italy}

\affiliation
{$^4$ 
Ru{d\llap{\raise 1.22ex\hbox
  {\vrule height 0.09ex width 0.2em}}\rlap{\raise 1.22ex\hbox
  {\vrule height 0.09ex width 0.06em}}}er
  Bo\v{s}kovi\'{c} Institute, HR-10$\,$001 Zagreb, Croatia}

\affiliation
{$^5$ Horia Hulubei National Institute of Physics and Nuclear Engineering,
  077125, Bucharest-Magurele, Romania}

\affiliation
{$^6$ Laboratoire Pluridisciplinaire Hubert Curien,  
CNRS-IN2P3/ULP, F-67$\,$037 Strasbourg, France}

\affiliation
{$^7$ Istituto Nazionale di Fisica Nucleare, Sezione di Napoli, 
 I-80126 Napoli, Italy}

\date{\today}
\begin{abstract}

 Multinucleon transfer reactions in $^{40}$Ca$+^{96}$Zr and 
 $^{90}$Zr$+^{208}$Pb have been measured at energies close to
 the Coulomb barrier 
 in a high resolution $\gamma$-particle 
 coincidence experiment. 
 The large solid angle magnetic spectrometer PRISMA coupled to the 
 CLARA $\gamma$-array has been employed.  
 Trajectory reconstruction has been applied for the complete 
 identification of transfer products.   
 Mass and charge yields, total kinetic energy losses, $\gamma$ 
 transitions of the binary reaction partners, and comparison of data with 
 semiclassical calculations are reported. 
 Specific transitions in $^{95}$Zr populated in one particle
 transfer channels are discussed in terms of particle-phonon couplings. 
 The $\gamma$ decays from states in $^{42}$Ca in the excitation
 energy region expected from pairing vibrations are also observed.

\end{abstract}


\maketitle

\section{Introduction}

 Our understanding of the structure of nuclei
 greatly benefited from reactions induced by a very 
 large variety of probes. With photons, electrons, mesons and nucleons  
 we learned about the properties of nuclei 
 in the vicinity of their ground states. These results were essential 
 for the formulation of nuclear models in terms of elementary modes like 
 collective surface vibrations and single particle degrees of freedom.
 Transfer reactions induced by light ions, deuteron, tritium and alphas 
 played an essential role for our knowledge of 
 particle-particle correlations 
 and led to the recognition of new elementary modes like 
 pair-vibration and pair-rotation \cite{Bro}.
 
 The acceleration of heavy ions offered the possibility to bring 
 together two complex systems. In the collision process they might 
 exchange several quanta, of energy and angular momentum and of 
 mass and charge, 
 or they could fuse giving rise to compound system 
 with very large excitation energy and intrinsic 
 angular momentum \cite{aagebook}. 
 By using heavy-ion fusion reactions
 the study of the nuclear structure has been extended
 in regions of very high angular momentum and large excitation energies. 
 This field of research turned out to be very prolific and for the last
 thirty years dominated the nuclear structure studies.  
 However, other aspects of the nuclear behavior \cite{wu} 
 remain largely unexplored. For instance the existence of the two 
 octupole phonons in $^{208}$Pb is still debated \cite{Schr,Woll,Vetter}, 
 and only recently
 the particle-vibration coupling scheme
 received new attention from the observation of large survival probabilities 
 of the octupole vibration in all neighboring
 nuclei populated in deep-inelastic or multi-nucleon transfer
 reactions with lead \cite{Rej}. 

 The identification of elementary modes of excitation in nuclei 
 turned out to be important also for the development of models 
 for the reaction.
 While it is natural to use the semi-classical approximation
 for the description of the relative motion, the coherent 
 excitation of the elementary modes (multi-phonons) 
 allows the use of the same semi-classical approximation 
 also for the treatment of the intrinsic degrees of freedom and
 thus to develop models  
 that are able to treat on the same footing phonons and 
 single particle degrees of freedom \cite{Win1,Vig}. 
 
 Heavy-ion reactions offer, in principle, an ideal tool for the study
 of the residual interaction in nuclei,  
 in particular the components responsible
 for the couplings between the phonon degrees of freedom and the single
 particle (particle-vibration coupling), and via  
 multi-nucleon transfer reactions the component responsible
 for particle correlations like the pairing interaction.
 However, the analysis and subsequent interpretation of these reactions
 turned out to be quite complex being the 
 information about correlations buried in the inclusive 
 character of the extracted cross sections \cite{Cor1,szi-capb}. 
 With the ability to measure
 individual transitions \cite{CaSn}, 
 a deeper insight into particle correlations 
 can be achieved. The new generation of large solid angle 
 spectrometers \cite{Stef,Vamos} coupled with
 $\gamma$ arrays \cite{Gad,Exo} gives the possibility  to look at 
 individual transitions, their population pattern and decay modes 
 via particle-$\gamma$ coincidences.
 The need to reach good resolution in all mass regions and to 
 measure low cross sections for massive transfer channels, 
 lead to the construction of PRISMA+CLARA. 
 In the magnetic spectrometer PRISMA \cite{Stef} reaction products 
 are identified via an event-by-event reconstruction of the ion trajectory
 inside the magnetic elements. Coincident $\gamma$ rays are 
 detected with CLARA \cite{Gad}, 
 where its high granularity allows to perform precise Doppler correction of
 transitions tagged with the spectrometer.  

 In this manuscript the multinucleon transfer reactions 
 in $^{40}$Ca$+^{96}$Zr and $^{90}$Zr$+^{208}$Pb, 
 studied at energies close to the Coulomb barrier with the PRISMA+CLARA 
 set-up are presented.  
 Both projectile and target are well known closed-shell nuclei 
 and therefore optimum candidates for having clean experimental and 
 theoretical conditions.

\section{The experimental set-up}

 Beams of $^{40}$Ca and $^{90}$Zr have been accelerated 
 on $^{96}$Zr and $^{208}$Pb targets, respectively, 
 by the Tandem and 
 ALPI booster of LNL at bombarding energies of 152 MeV and 560 MeV
 \cite{fusion06}, respectively. 
 The beams had an average intensity of $\simeq$ 3pnA.
 The $^{96}$Zr (in form of oxide) and $^{208}$Pb targets 
 had a thickness of 150 and 290 $\mu$g/cm$^2$, respectively.   
 Both targets had an isotopic enrichment of 99.9\% and consisted of a 
 strip of 1 mm on a 20 $\mu$g/cm$^2$ C-layer.    

\begin{figure}[thb]
\vspace{6mm} 
\centering
%\includegraphics[width=0.405\textwidth]
%{fig/fig1.ps} \vspace{-0.5mm}
%{szilner-fig1.ps} \vspace{-0.5mm}
 \caption{Top: Example of two-dimensional $\Delta E-E$ matrix corresponding 
          to one 
          section of the IC, obtained in the reaction $^{90}$Zr$+^{208}$Pb.
          The most intense band corresponds to Z=40.   
          The diagonal line is a trivial effect of detector thresholds.
          The matrix has been constructed by requiring the coincidence 
          with at least 
          one $\gamma$-ray in CLARA, so the contribution 
          of elastic scattering is suppressed.
          Bottom: Two-dimensional Range vs total energy matrix   
          obtained in the reaction $^{40}$Ca$+^{96}$Zr, where all 
          sections of the IC have been included. 
          The most intense band corresponds to Z=20.
          The two bands located at the bottom of the matrix correspond to 
          $^{12}$C and $^{16}$O contaminants present in the target.  
}
\label{dE-E}
\end{figure}
     
 Projectile-like products have been selected with the 
 magnetic spectrometer PRISMA placed at $\theta_{\rm lab}$=68$^{\circ}$ 
 for the $^{40}$Ca$+^{96}$Zr and $\theta_{\rm lab}$=61$^{\circ}$ for 
 the $^{90}$Zr$+^{208}$Pb reactions.  
 PRISMA consists of a magnetic quadrupole singlet, 
 placed at 50 cm from the target, 
 and a magnetic dipole (60$^{\circ}$ bending angle and 1.2 m curvature radius).
 Its main characteristics are the large 
 solid angle of $\simeq$ 80 msr (corresponding to 
 $\pm 6^{\circ}$ in $\theta$ and 
 $\pm 11^{\circ}$ in $\phi$), a momentum acceptance 
 $\Delta p/p = \pm 10\%$ and 
 a dispersion of $\simeq$ 4 cm per percent in momentum. 
 A two-dimensional position sensitive micro-channel plate (MCP) detector
 \cite{Mon} is placed at the entrance of the spectrometer providing 
 a start signal for time-of-flight measurement with sub-nanosecond resolution 
 and $X_{\rm i}$,$Y_{\rm i}$ signals with 1 mm resolutions. 
 Ions pass through the optical elements of the spectrometer and after a path  
 of $\simeq$ 6.5 m, enter a focal plane detector \cite{Beg}. 
 This is made of a parallel plate of multiwire-type (MWPPAC) divided into ten 
 sections, providing timing and $X_{\rm f}$,$Y_{\rm f}$ position 
 signals derived via a 
 delay line method, with resolutions similar to the MCP ones.  
 Behind the MWPPAC an array of a transverse field 
 multiparametric ionization chambers (IC) is placed, providing nuclear charge 
 ($\Delta E$) and total energy ($E$) (with a $\sim 2-3\%$ resolution).   
 The IC is divided into ten transversal sections (like the MWPPAC) with four 
 $\Delta E$ subsections each. Combining the $\Delta E$ subsections 
 and adjusting properly the gas pressure, one can stop inside the IC ions  
 that differ by more than 
 20$\%$ in kinetic energy, and optimize the $Z$ and total energy resolutions.

 The $\gamma$-ray array CLARA \cite{Gad} consists of 24 HP-Ge 
 clover-type detectors 
 placed to form a 2$\pi$ 
 configuration close to the target position and opposite to PRISMA. 
 Each clover detector is composed of four crystals mounted 
 in a single cryostat 
 and surrounded by an anti-Compton shield, ensuring a peak-to-total ratio of 
 $\simeq50\%$. The total photopeak efficiency of CLARA is of the order 
 of 3$\%$ for $E_{\gamma}$=1.33 MeV.    
 Typical $\gamma$-ray energy resolutions obtained after Doppler correction  
 are 0.6$\%$ to 0.9$\%$ 
 over the whole velocity distribution of 
 the projectile-like products detected in PRISMA.

\section{PRISMA data analysis}

 As it is well known, a conventional magnetic spectrometer provides 
 the ratio of the momentum over the atomic charge state. To obtain  
 the mass, an additional parameter is needed, and a commonly used one is the 
 time-of-flight.  
 With very large solid angle spectrometers, like PRISMA, 
 the mass identification of the reaction products can only be obtained
 via an event-by-event reconstruction of the ion trajectory inside the 
 magnetic elements.  
 The reconstruction of the trajectory is here obtained from the measurement
 of entrance detector positions $X_{\rm i}$,$Y_{\rm i}$, focal plane positions 
 $X_{\rm f}$,$Y_{\rm f}$
 and time-of-flight $\tau_{_{\rm TOF}}$. 
 For the reconstruction a following fast algorithm has been employed.
 Taking advantage of the very large longitudinal dimension of PRISMA  
 (6.5 m) with respect to the transversal one (0.2 m) and considering that 
 the fringing fields 
 can be neglected because of the large dimensions of the magnetic elements,  
 the trajectory may be assumed to be planar after the quadrupole. 
 These assumptions, planarity of the trajectory 
 and the weak effect of the fringing fields,  
 have been carefully checked with a simulation \cite{latina} of the ion
 transport
 through the spectrometer by using the actual 
 fields and by mimicking the distribution of the 
 reaction products with a Monte Carlo method. The trajectories are uniquely 
 determined by two parameters, the bending radius in the dipole 
 and the ratio of the quadrupole and dipole magnetic fields. 
 Being the magnetic fields known, the bending radius $\rho$ remains the
 only parameter 
 to be searched for.  

\begin{figure}%[tbh]
\vspace{6mm} 
\centering
\includegraphics[width=0.405\textwidth]
%{fig/fig2.ps} \vspace{-0.5mm}
{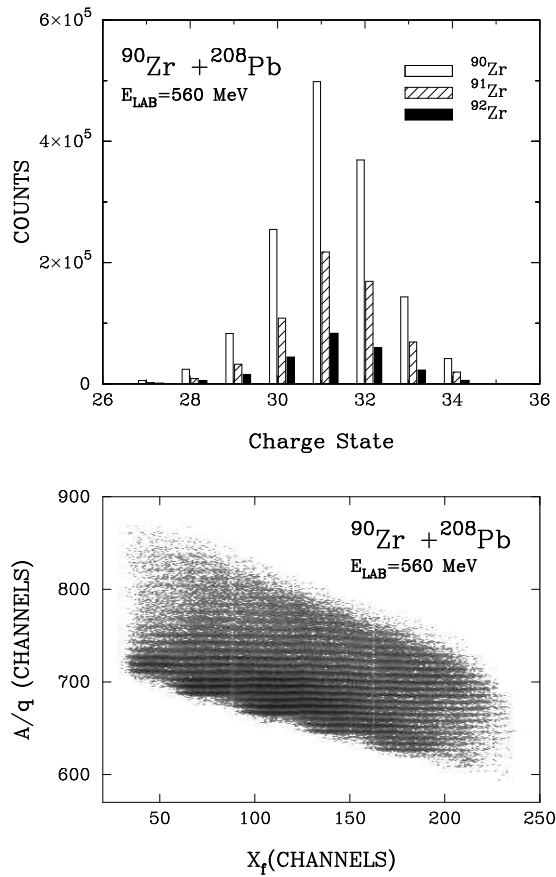} \vspace{-0.5mm}
 \caption{Top: Atomic charge state distribution for $^{90,91,92}$Zr 
          ions selected with PRISMA 
          in the reaction $^{90}$Zr$+^{208}$Pb.
          Bottom: Mass over atomic charge state $A/q$ vs
          horizontal MWPPAC position $X_f$.}
\label{charge-mass-x}
\end{figure}

 We briefly describe below the procedure adopted to identify the mass $A$ 
 and the nuclear charge $Z$ of the reaction products. 
 The identification of nuclear charge $Z$ is obtained 
 through the measurement of energy loss $\Delta E$ in the IC, which 
 provides also the total energy $E$. 
 Fig. \ref{dE-E} (top panel) displays an example of $\Delta E-E$ matrix 
 for the $^{90}$Zr$+^{208}$Pb system, where only one of the 
 central sections of the IC is considered. Here a clear separation of the 
 different Z is obtained. 
 We have to keep in mind that ions reach the IC with a broad range of
 kinetic energies 
 and directions, thus to get the desired Z resolution for 
 all the reaction products reaching the IC, one has to properly take into 
 account the direction followed by the 
 different ions. 
 One can estimate the path (Range) from the signal of  
 each subsection and the position information 
 of MCP and MWPPAC detectors. 
 An example of Range vs $E$ matrix is displayed in the same figure 
 (bottom panel) for 
 the $^{40}$Ca$+^{96}$Zr reaction. 
 The most intense bands in Fig. \ref{dE-E} 
 correspond to the Z of the entrance channel for both systems. 
 As pointed out before mass identification of the reaction products 
 requires the determination of their trajectory in the apparatus. 
 From this trajectory we obtain the bending radius $\rho$ in the dipole 
 and the total length $L$ up to the MWPPAC. 
 Thus, first we obtain the quantity: 
   
\begin{equation}
\frac{B ~ \rho ~ \tau_{_{\rm TOF}}} {L}\Rightarrow \frac{A}{q}
\label{for:Aq}
\end{equation}
 
\noindent
 that is proportional to the ratio $A/q$. 
 By plotting this quantity as a function of the $X_f$ position in the MWPPAC 
 a clear discrimination with characteristic repetitive pattern
 of the different $A/q$ 
 is obtained (see bottom panel of Fig. \ref{charge-mass-x}). 
 The large acceptance of the spectrometer is reflected 
 in the fact that different atomic charge states cover several sections
 in the focal plane, which are all joined together in the above 
 mentioned figure.  
 
 The actual mass is finally obtained by constructing the quantity : 

\begin{equation}
\frac{E ~ \tau_{_{\rm TOF}}}{B ~ \rho ~ L} \Rightarrow q 
\label{for:q}
\end{equation} 

\noindent
 that is proportional to the atomic charge state $q$. 
 In the above equations $B$ is the strength of the dipole field 
 and $E$ the kinetic energy of the ion.  

\begin{figure}
\vspace{6mm} 
\centering
\includegraphics[width=0.405\textwidth]
%{fig/fig3.ps} \vspace{-0.5mm}
{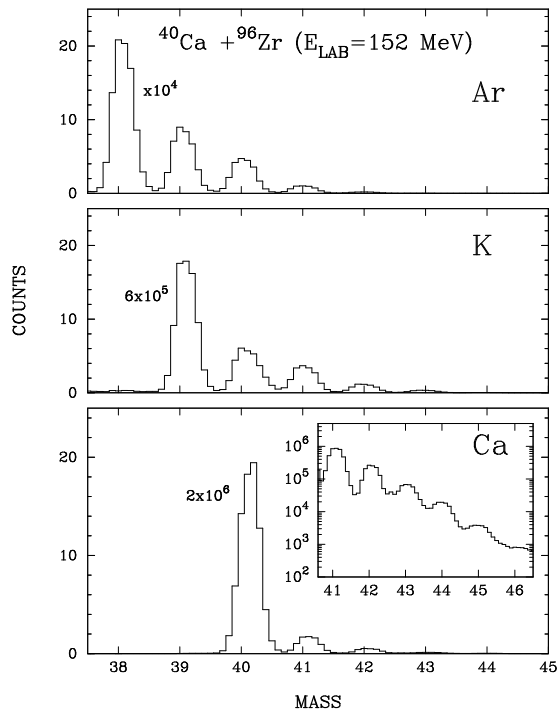} \vspace{-0.5mm}
 \caption{Mass spectra for argon, potassium and calcium isotopes 
  populated in the $^{40}$Ca$+^{96}$Zr reaction (note the different 
  scaling factors in each panel). The inset 
  in the bottom panel displays the neutron pick-up channels in 
  logarithmic scale.    
  }
\label{mass-neutron}
\end{figure}

 The top panel of Fig. \ref{charge-mass-x} shows, for 
 the $^{90}$Zr$+^{208}$Pb reaction, 
 the extracted atomic charge state distribution for different 
 zirconium isotopes.
 The broad distribution, characteristic of heavy ions,  
 does not depend on the isotope, and have a centroid that matches 
 the one calculated according to Ref. \cite{shima}. 
 The reconstructed mass spectra 
 for calcium, potassium, and argon isotopes are shown 
 in  Fig. \ref{mass-neutron} 
 for the $^{40}$Ca$+^{96}$Zr reaction. 

\begin{figure}[h]
\vspace{6mm} 
\centering
\includegraphics[width=0.420\textwidth]
%{fig/fig4.ps} \vspace{-0.5mm}
{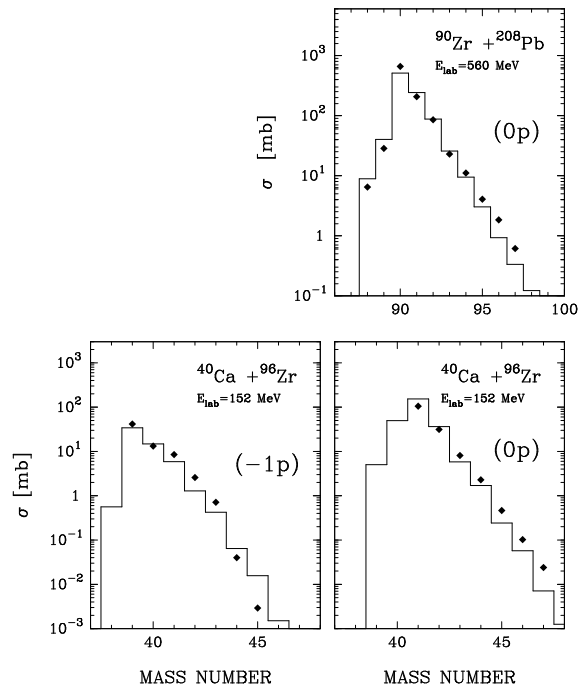} \vspace{-0.5mm}
 \caption{Top: Total cross sections for pure neutron pick-up channels in the 
  $^{90}$Zr$+^{208}$Pb reaction. 
  Bottom: Total cross sections for pure neutron pick-up (right panel) and 
  one-proton stripping (left panel) channels in the $^{40}$Ca$+^{96}$Zr
  reaction. 
  The points are the experimental data and the histograms 
  are the calculation performed with the code GRAZING.  
}
\label{sigma-CaZrPb}
\end{figure}

 From the grazing character of these reactions 
 we know that most of the yield in the different transfer channels is
 concentrated 
 in a narrow angular range close to the grazing with a shape of the angular 
 distribution weakly dependent on the isotope (cfr. Refs. \cite{Cor1,szi-capb}
 and references 
 therein for illustration). 
 Exploiting the large angular acceptance of PRISMA one obtains
 a reasonable estimation  
 of the relative production yield. 
 The total yields for the pure neutron transfer channels 
 are shown in Fig. \ref{sigma-CaZrPb} for the two reactions,  
 together with the yield of the one proton stripping channel 
 for the $^{40}$Ca$+^{96}$Zr system. 
 In the same figure the results of a semiclassical calculation 
 \cite{Win1,grazing_prog} are shown. 
 The experimental yields have been normalized to the computed +1n channel 
 and the same normalization constant has been kept for other neutron 
 pick-up and 
 the proton stripping channels. 
 The good agreement between experiment and theory gives us confidence
 on the correct 
 procedure adopted for the trajectory reconstruction and of the good behavior 
 of the device.   
 
 The dependence of the cross sections on the number of transferred 
 neutrons is very similar to the one observed in other studied systems 
 \cite{Rehm1,Cor1,szi-capb}. 
 We remind that the shape of the proton stripping channels illustrates 
 the independence between the proton and neutron transfer degrees of freedom. 
 The apparatus allowed the identification of neutron stripping channels 
 which are, for stable beams, 
 strongly suppressed by optimum $Q$-value considerations \cite{dasso}. 
 Since these channels may receive 
 considerable contributions from evaporation processes,
 these effects have to be 
 included in the theoretical model. The calculated cross sections 
 have been obtained by using the semiclassical model GRAZING 
 \cite{Win1,grazing_prog}. 
 This model calculates the evolution of the reaction
 by taking into account, besides 
 the relative motion variables, the intrinsic degrees of
 freedom of projectile and target.  
 These are the isoscalar surface modes and the single nucleon transfer
 channels. 
 The multinucleon transfer channels are described via a multi-step mechanism.
 The relative motion of the system is calculated in a nuclear
 plus Coulomb field 
 where for the nuclear part the empirical potential of 
 Ref. \cite{aagebook} has been used. 
 The excitation of the intrinsic degrees of freedom is obtained by
 employing the 
 well known form factors for the collective surface vibrations
 and the one-particle 
 transfer channels \cite{abs1,abs2}. 
 The model takes into account in a simple way the effect of
 neutron evaporation.

\section{PRISMA+CLARA data analysis}

\begin{figure}[h]
\vspace{6mm} 
\centering
\includegraphics[width=0.425\textwidth]
%{fig/fig5.ps} \vspace{-0.5mm}
{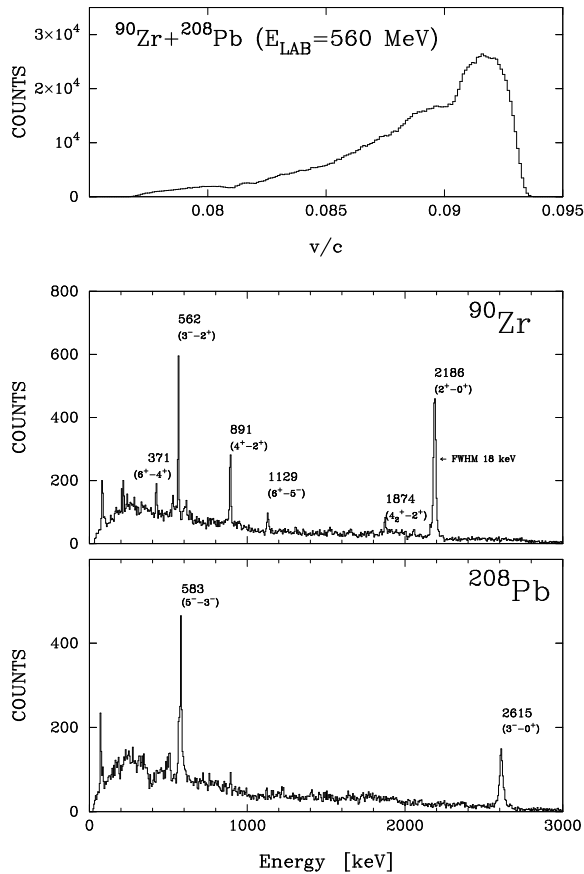} \vspace{-0.5mm}
 \caption{Top: Reconstructed velocity distribution for the $^{90}$Zr ions 
 in the $^{90}$Zr$+^{208}$Pb reaction. 
 Middle : Doppler corrected $\gamma$-ray spectrum for $^{90}$Zr. 
 Bottom: Doppler corrected $\gamma$-ray spectrum for $^{208}$Pb, 
 obtained assuming a binary character of the reaction.   
 The $\gamma$-ray spectra have been obtained by subtracting the wrongly
 Doppler corrected contribution of the binary partner.
 Main $\gamma$-ray transitions are marked with their energy (in keV) 
 together with the spin and parity of the connected states. 
}
\label{velocity-dist}
\end{figure}

 The particle-$\gamma$ coincidence obtained from the coupling of CLARA with 
 PRISMA allows to attribute to each specific reaction product its 
 characteristic $\gamma$ rays. 
 Since the $\gamma$ rays are emitted in-flight it is mandatory to perform 
 Doppler correction. This is done from the knowledge of the trajectory 
 reconstructed in PRISMA, providing the velocity vector 
 of the emitting nuclei. 
 The top panel of Fig. \ref{velocity-dist} shows a reconstructed
 velocity distribution 
 for the $^{90}$Zr ions in the $^{90}$Zr$+^{208}$Pb reaction. 
 The middle panel depicts the corresponding Doppler corrected 
 $\gamma$-ray spectrum, illustrating the good resolution obtained with 
 such a procedure (the 2$^+\rightarrow 0^+$ transition at 2186 keV has a 
 FWHM of 18 keV).
 The characteristic $\gamma$ rays of the undetected heavy partner can be 
 Doppler corrected by assuming a binary character of the reaction.  
 In the bottom panel of Fig. \ref{velocity-dist} 
 we display the $\gamma$-ray spectrum for $^{208}$Pb, where we clearly observe
 the decay of the $3^-$ and $5^-$ states. 

\begin{figure}[h]
\vspace{6mm}
\centering
\includegraphics[width=0.405\textwidth]
%{fig/fig6.ps} \vspace{-0.5mm}
{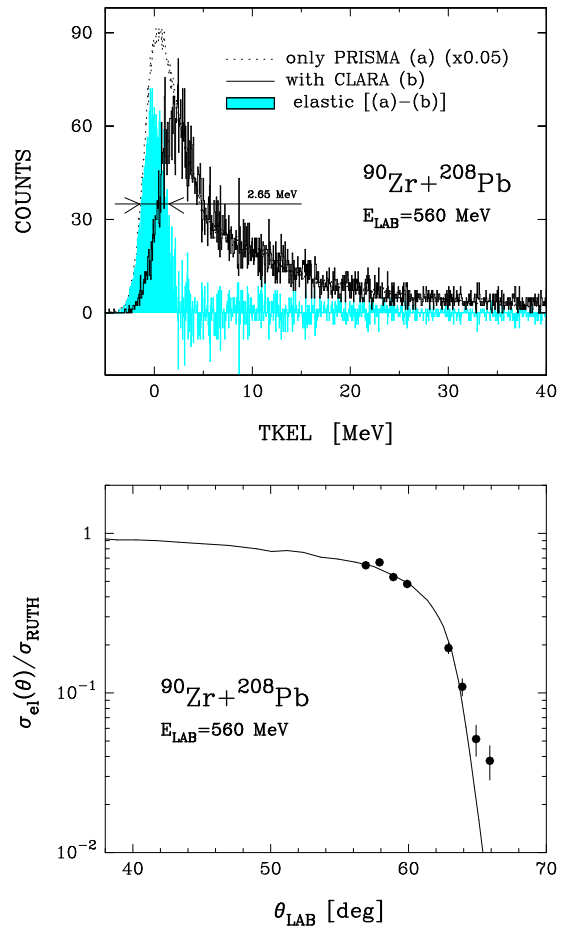} \vspace{-0.5mm}
 \caption{(Color online) Top: Experimental angle integrated total kinetic energy loss 
 distributions
 (TKEL) for $^{90}$Zr in the $^{90}$Zr$+^{208}$Pb reaction 
 (a) without coincidence with $\gamma$ rays and (b)
 with at least one $\gamma$ ray detected in CLARA. 
 The two spectra are normalized in such a way
 that the high TKEL tails match. 
 The gray area corresponds to the subtraction between the two 
 spectra [(a)-(b)], 
 giving a peak whose width is $\sim$ 2.65 MeV.  
 Bottom: Experimental (points) and GRAZING calculated (curve)
 differential cross section for elastic scattering, normalized to Rutherford. 
 Only statistical errors are included.}
\label{subtraction}
\end{figure}

 The present set-up offers the possibility to separate elastic from inelastic 
 scattering. 
 The pure elastic scattering is here determined by comparing the
 events with and without $\gamma$ coincidences.  
 In the top panel of Fig. \ref{subtraction} are shown the total kinetic 
 energy loss (TKEL) 
 spectra for $^{90}$Zr with and without $\gamma$-coincidence, 
 normalized in the tail (large TKEL) region.  
 By subtraction, we obtain the contribution of pure elastic. 
 This subtracted spectrum is characterized by a narrow peak centered at 
 TKEL $\simeq$ 0 MeV with a FWHM of 2.65 MeV. Moreover, its centroid is 
 separated 
 by 2.15 MeV from the maximum of the TKEL spectrum in coincidence with CLARA, 
 whose value is very close to the inelastic excitation of the 
 first 2$^+$ state in $^{90}$Zr. 
 Such a procedure should be reliable, 
 provided that the shape of the spectrum in coincidence 
 with $\gamma$ rays only weakly depends on the $\gamma$ multiplicity. 
 This fact is fulfilled for nuclei having low level density close to the 
 ground state and rather narrow ($\simeq$ 2-3 MeV) TKEL distributions, as 
 for the present near closed-shell nuclei.   

 By repeating this subtraction in steps of one degree over the entrance 
 angular range ($\Delta \theta_{\rm lab} = 12^{\circ}$) of PRISMA one obtains 
 the elastic angular distribution whose ratio to Rutherford is shown
 in the bottom panel of Fig. \ref{subtraction}, 
 in comparison with the results of GRAZING calculations \cite{grazing_prog}. 
 The very pronounced fall-off of the elastic cross section for 
 large angles clearly
 indicates that the elastic scattering for this system 
 is dominated by strong absorption.  
 The good agreement between theory and experiment gives us confidence 
 on the used potential and on the fact that the included reaction channels 
 correctly describe the depopulation of the entrance channel (absorption).

\begin{figure}[h]
\vspace{6mm} 
\centering
\includegraphics[width=0.405\textwidth]
 % {fig/fig7.ps} \vspace{-0.5mm}
 {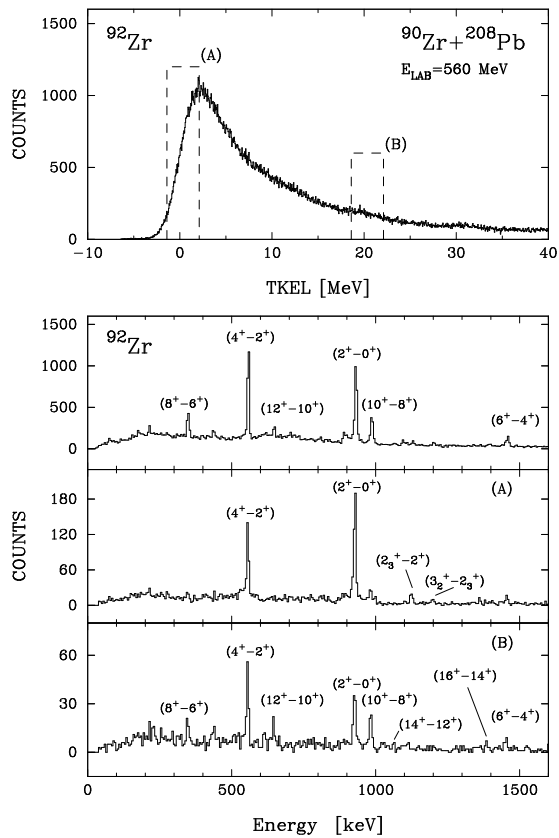} \vspace{-0.5mm}
 \caption{Top: TKEL distribution for $^{92}$Zr produced in the 
  $^{90}$Zr$+^{208}$Pb reaction. Bottom: associated $\gamma$-ray spectra 
  for $^{92}$Zr without conditions on TKEL (top) and 
  conditioned (middle and bottom) 
  with different regions of TKEL distributions, marked as (A) and (B) 
  in the TKEL spectrum and with gates $\simeq$ 3 MeV wide.}
 \label{qvaluegamma}
\end{figure}

 Grazing reactions populate regions of excitation energy and spin differing  
 substantially from the ones reached by fusion evaporation reactions. 
 This can be here checked by a careful correlation 
 of the TKEL spectra obtained with PRISMA with the coincident 
 $\gamma$ rays of CLARA. 
 The top panel of Fig. \ref{qvaluegamma} shows the 
 TKEL spectrum for $^{92}$Zr from the $^{90}$Zr$+^{208}$Pb reaction, 
 while the bottom panel shows the corresponding $\gamma$-ray spectra 
 obtained without (top panel) and with (middle and bottom panels) different 
 conditions on the TKEL. 
 States up to spin 16$^+$ and 
 excitation energy of about 7.5 MeV, have been observed. 
 By gating on the low TKEL  
 region (A) the two lowest yrast transitions dominate the spectrum.  
 By gating instead on the TKEL region around 20 MeV (B)   
 the spectrum displays transitions coming from the decay of high-spin states 
 (we remark that the excitation energy of both light and heavy
 fragments is embedded into the TKEL distribution).

\begin{figure}[h]
\vspace{6mm} 
\centering
\includegraphics[width=0.430\textwidth]
 % {fig/fig8.ps} \vspace{-0.5mm}
 {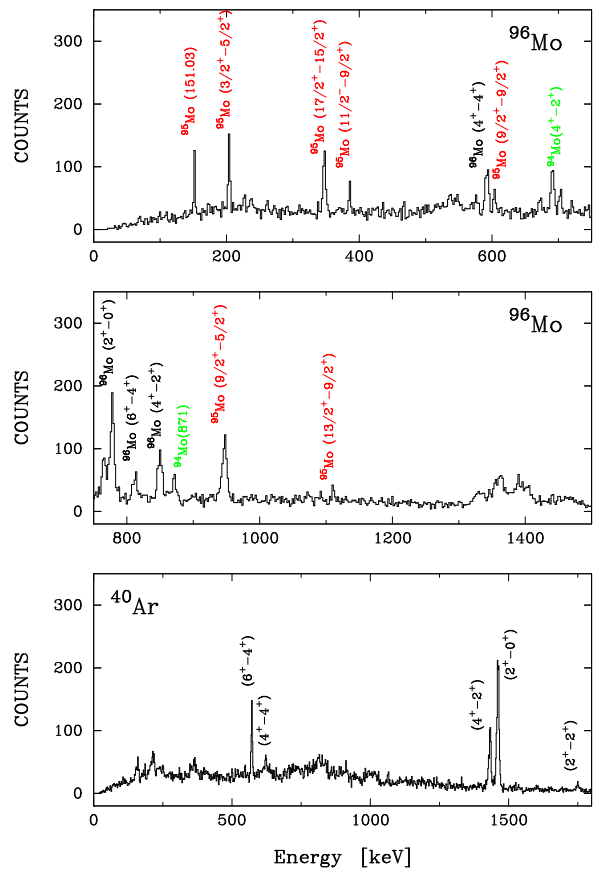} \vspace{-0.5mm}
 \caption{(Color online) $\gamma$ spectra for the $-2p+2n$ channel in the reaction 
  $^{40}$Ca$+^{96}$Zr Doppler corrected for the heavy (top two frames)
  and light fragments (bottom frame). 
  To have a better identification of the different $\gamma$ lines 
  for the heavy fragment we used an expanded energy scale.}
 \label{gammaLH}
\end{figure}

 Exploiting the binary character of the studied reaction  
 one can investigate the final mass partition influenced 
 by evaporation processes. 
 Gating with PRISMA on a specific $Z$ and $A$ (light partner)
 the velocity vector of the undetected heavy partner can be evaluated  
 and applied for the Doppler correction of its 
 corresponding $\gamma$ rays. In those spectra not only 
 the $\gamma$ rays belonging to the primary binary partner are present   
 but also the ones of the nuclei produced after evaporation takes place.   
 For $^{40}$Ar detected with PRISMA in the reaction 
 $^{40}$Ca$+^{96}$Zr, we show in Fig. \ref{gammaLH} 
 the $\gamma$ spectra of the light and heavy partners.  
 For this $-2p+2n$ channel, about 60$\%$ of the yield 
 corresponds to $^{96}$Mo, while the rest 
 is equally shared between isotopes corresponding to the evaporation 
 of one and two neutrons, leading to $^{95}$Mo and $^{94}$Mo, respectively. 
 For even weaker channels like the $-4p$ channel ($^{36}$S) 
 the $\gamma$ rays belonging to the primary binary 
 partner ($^{100}$Ru) have negligible yield. 
 On the other hand, for the $+2n$ channel ($^{42}$Ca) 
 more than 90$\%$ of the yield corresponds to the true binary partner, 
 i.e. $^{94}$Zr. 
 This behavior is closely connected with the observed TKEL. 
 For the neutron pick-up channels the major contribution 
 in the TKEL is close to the optimum $Q$ values ($Q_{\rm opt} \simeq 0$), 
 while in the proton stripping channels larger TKEL are observed,  
 thus the neutron evaporation has a stronger effect on the 
 final mass partition. 
 The importance of neutron evaporation in the modification of 
 the final yield distribution was outlined  
 in inclusive measurements \cite{Cor1,szi-capb}. 
 A direct signature of this effect was observed 
 by correlating projectile-like and target-like 
 fragment isotopic yields via $\gamma-\gamma$ coincidences 
 \cite{Asztalos,Broda}.

\section{On the structure of $^{95}$Zr and $^{42}$Ca} 

 The experimental yields have been interpreted with 
 a model that explicitly treats the internal degrees of freedom of the
 two ions in terms of elementary modes, surface vibration and single particles.
 The excitation and transfer processes are  mediated by  
 the well known single-particle form factors for the fermion degrees 
 of freedom and by the collective form factors, nuclear plus coulomb, for the
 vibrational modes. Following this description heavy ion collisions
 provide a suitable tool for the studies of the particle-vibration
 coupling scheme, in fact it is through the excitation of these elementary
 modes that energy  and angular momentum are transferred from the relative 
 motion to the intrinsic degrees of freedom and that mass and charge are 
 exchanged among the two parters of the collision. 
 We remind that the  
 de-excitation spectra of the produced isotopes may quite differ 
 from the excitation ones. 
 The quadrupole and octupole matrix elements generally play
 a major role in the excitation process, while the de-excitation 
 may be dominated by very small components in the wave function.

 The reactions we are analyzing in this manuscript are well suited for these 
 studies since they involve magic and semi-magic nuclei. 
 The spectra of the neighboring nuclei,
 populated in the reaction, comprise partly single particle or single hole
 states and partly states that involve combinations of single-particle or 
 hole with a collective boson. 
 Here we concentrate on spectra of 
 $^{95}$Zr populated via one-neutron pick-up reaction in the
 $^{40}$Ca$+^{96}$Zr collision.  
 Our analysis follows 
 the one of Ref. \cite{Kadi} where the particle-phonon states in  
 $^{209}$Bi and $^{207}$Pb, populated in deep-inelastic heavy ion collision, 
 were analyzed. 
 
\begin{figure}[hbt]
\vspace{6mm}
\centering
\includegraphics[width=0.45\textwidth]
%{fig/fig9.ps} \vspace{-0.5mm}
{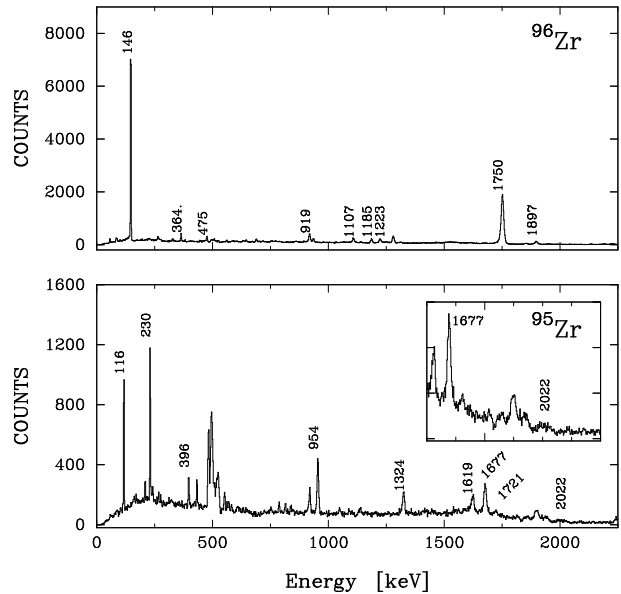} \vspace{-0.5mm}
 \caption{Gamma-ray spectra for $^{96}$Zr (top) and $^{95}$Zr (bottom) 
 obtained in the $^{40}$Ca$+^{96}$Zr reaction (see text).
}
\label{gammaZrNb}
\end{figure}

 The $^{208}$Pb nucleus constitutes the ideal laboratory for the study of 
 particle-phonon states, being its first excited state  
 the collective 3$^-$
 at 2.62 MeV and its first positive parity state the 2$^+$ at 4.05 MeV.
 The collectivity of the 3$^-$ ($B(E3; 3^- \rightarrow 0^+)$= 34 W.u.) 
 is not dominated 
 by few particle-hole components but derives from the cooperative
 action of many configurations.
 The $^{96}$Zr nucleus presents a more complicated situation, its
 low-energy spectra is dominated by a 2$^+$ state at 1.75 MeV and by
 the 3$^-$ at 1.90 MeV.  The last state 
 is very collective ($B(E3; 3^- \rightarrow 0^+)$ = 51 W.u.) 
 and decays via an E1 transition 
 to the 2$^+$ and via an E3 transition to the ground state \cite{nds96}.
 The $\gamma$ spectra for $^{96}$Zr and $^{95}$Zr are shown 
 in Fig. \ref{gammaZrNb}, they  
 have been obtained by applying Doppler correction for  
 target-like partners corresponding to $^{40}$Ca and $^{41}$Ca 
 detected in PRISMA, respectively. 

 The ground state of $^{95}$Zr is well described by a neutron-hole
 in the $d_{5/2}$ orbital. By coupling this hole state with the 3$^-$ one
 expects a sextuplet of states (1/2$^-$,3/2$^-$, ...,11/2$^-$) 
 at an energy close to the one of the 3$^-$. 
 Similarly, by coupling  
 the same hole state
 to the first 2$^+$ one expects a quintuplet (1/2$^+$, 3/2$^+$ ...,9/2$^+$) 
 at an energy close to the one of the 2$^+$ state in $^{96}$Zr. 
 The reaction mechanism does not
 populate the components of the two multiplets uniformly 
 but favors the stretched configurations
 11/2$^-$ and 9/2$^+$ since the transfer probability has its maximum 
 at the largest angular momentum transfer.
 From the adopted levels of $^{95}$Zr, the state at 2025 keV 
 (2025$\pm$7 keV) 
 populated in $(p,d)$ and $(^3{\rm He},\alpha)$ reactions 
 (recognized as 9/2$^-$, 11/2$^-$)
 \cite{Bing} is a natural candidate for the
 stretched configuration $|3^-,(d_{_{5/2}})^{-1}>$. 
 Very recently, the level scheme of $^{95}$Zr has been 
 re-measured \cite{Pantelica} in heavy-ion 
 induced fission reactions and the sequence of $\gamma$ rays of 
 229.7 [11/2$^- \rightarrow$ (9/2$^+$)], 115.8 [(9/2$^+) \rightarrow$
 (7/2$^+$)] and 1676.8 [(7/2$^+$) $\rightarrow$ 5/2$^+_{g.s.}$] keV     
 has been proposed for the decay of the 11/2$^-$ state.
 In Fig. \ref{gammaZrNb} the measured spectrum of $^{95}$Zr shows
 very clearly all these transitions. In addition, 
 a new transition at E$_{\gamma}$=2022 keV is visible 
 (the width of the 2022 keV peak is partially affected by
 the wrongly Doppler corrected 3/2$^+ \rightarrow
 7/2^-_{g.s.}$ transition from $^{41}$Ca), 
 that we naturally interpret as the E3 decay of the 11/2$^-$ 
 to the ground state. 
 The intensity of this transition, relative to the E1, 
 is very similar to the one observed in $^{96}$Zr 
 thus reinforcing our interpretation that the 11/2$^-$ is a member 
 of a boson-hole $|3^-,(d_{_{5/2}})^{-1}>$ multiplet.

\begin{figure}[h]
\vspace{6mm}
\centering
\includegraphics[width=0.405\textwidth]
%{fig/fig10.ps} \vspace{-0.5mm}
{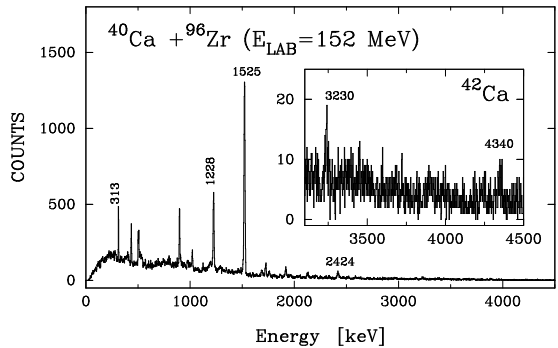} \vspace{-0.5mm}
 \caption{$\gamma$-ray spectrum with expanded region in inlet for
 $^{42}$Ca obtained in the
  $^{40}$Ca$+^{96}$Zr reaction.}
\label{gammaZr}
\end{figure}
 
 The closed-shell systems studied in this work are well suited for the 
 identification of states reached via the addition and/or the removal of 
 pairs of nucleons. Those states have been studied with light ion reactions 
 and formed the basis for the identification of pairing vibration 
 degrees of freedom in the nuclear medium \cite{Bro}. 
 In a previous experiment with $^{40}$Ca$+^{208}$Pb (cfr. \cite{szi-capb}
 and Refs. therein) we have noticed that in order to obtain a good description 
 of the experimental total cross sections for the different isotopes
 populated in the reaction one has to include, in the theoretical model,
 the degrees of freedom related to the transfer of pairs of nucleons, 
 both protons and neutrons. 
 These degrees of freedom are treated as pair-vibrational modes and for their
 excitation we have used the form factors as provided by the macroscopic model.
 The influence of these degrees of freedom is particularly visible in the
 proton sector since the cross sections of proton stripping channels are 
 much smaller than the neutron pick-up ones. 
 
 To have evidence of the excitation
 of these modes in the neutron sector 
 we have analyzed in detail \cite{szilner-epja} the 
 total kinetic energy loss (TKEL) spectra
 of $^{42}$Ca, the two neutron pick-up channel. Here most
 of the cross section is concentrated in
 a pronounced peak at an energy that is compatible with
 the excitation of a group of $0^+$ states at 
 $\sim$6 MeV where a pairing-vibrational state should be 
 located \cite{Bro}.

 The present set-up should allow the observation of the decay pattern of the 
 populated $0^+$ states.  In Fig. \ref{gammaZr} 
 we show the $\gamma$-spectrum for $^{42}$Ca obtained in the 
 reaction  $^{40}$Ca$+^{96}$Zr.
 We observe here (see expanded region) a $\gamma$-transition at 4340 keV 
 which is 
 consistent with a decay from a level at 5.8 MeV to the 2$^+_1$ state 
 \cite{tpca,will}. 
 The limited statistics accumulated for this transition (we  
 remark that such high energy $\gamma$ rays have a low photo-peak 
 efficiency) does not allow  to deduce the spin of the populated level, 
 though the distribution over the rings of CLARA shows an 
 isotropic pattern but with very large error-bars.  
 In the expanded $\gamma$ spectrum we also observe a $\gamma$ transition of
 3230 keV, which is the main branch of the decay from 
 the 2$^+$ state at 4760 keV, strongly populated in 
 $(t,p)$ reactions \cite{tpca,will}.

\section{Summary}

 Multi-nucleon transfer reactions have been studied with the
 large solid angle magnetic spectrometer PRISMA, where ions identification is 
 achieved by reconstructing event-by-event the trajectory of the ions 
 in the magnetic elements. Experimental yields 
 have been compared with semi-classical models
 that include surface vibrations and single particle transfer modes 
 and taking into account the effect of neutron evaporation. 
 By coupling the PRISMA spectrometer with the high resolution CLARA 
 $\gamma$-array  we showed how the transfer process populates  
 reaction products in 
 regions of angular momentum and excitation energy quite different
 from fusion-evaporation processes. This has been exploited to gain information
 on states that can be interpreted along the 
 particle-vibration coupling scheme. 
 In particular, we discussed the low energy spectrum of  $^{95}$Zr 
 populated in one-neutron pick-up
 reactions.
 Very preliminary results on the possible pairing vibration in $^{42}$Ca
 are also reported.

\subsubsection*{Acknowledgments}

The authors are grateful to the LNL Tandem-ALPI stuff for providing us with 
the good quality beams and the target laboratory for the excellent target 
preparation. This work was partly supported by the 
European Community FP6 -
Structuring the ERA - Integrated Infrastructure Initiative - contract
EURONS No. RII3-CT-2004-506065. 
This work was, also, supported in part by the Croatian Ministry of
Science, Education and Sports, Grant No. 0098-1191005-2890.

%%%%%%%%%%%%%%%%%%%%%%%%%%%%%%%%%%%
\end{document}